\documentclass[sigconf]{acmart}
\usepackage{times}  
\usepackage{helvet}  
\usepackage{courier}  
\usepackage{url}  
\usepackage{graphicx}  
\usepackage{tabularx}
\usepackage{amsmath}

\usepackage{amssymb}

\usepackage{amsfonts}
\usepackage{subcaption}
\usepackage{multirow}

\graphicspath{{images/}}

\setcopyright{none}

\newtheorem{problem}{Problem}

\acmConference[RecNLP 2019]{The AAAI 2019 Workshop on Recommender Systems and Natural Language Processing}{January 2019}{Honolulu, Hawaii, USA}
\copyrightyear{2019}

\begin{document}

\title[Learning Representations from Product Titles]{Learning Representations from Product Titles for Modeling Shopping Transactions}
\author{Binh Nguyen}
\affiliation{%
  \institution{Department of Informatics, SOKENDAI}
  \city{Tokyo}
  \state{Japan}
}
\email{binh@nii.ac.jp}

\author{Atsuhiro Takasu}
\affiliation{%
  \institution{National Institute of Informatics}
  \city{Tokyo}
  \state{Japan}
}
\email{takasu@nii.ac.jp}

\begin{abstract}
Shopping transaction analysis is significant in understanding the shopping behaviors of customers. Existing models such as association rules are poor at modeling products which have short purchase histories and cannot be applied to new products (the cold-start problem). In this paper, we propose BASTEXT, an efficient model of shopping baskets and the texts associated with the products (e.g., product titles). The model's goal is to learn the product representations from the textual contents, that can capture the relationships between the products in the baskets. Given the products already in a basket, a classifier identifies whether a potential product is relevant to the basket or not, based on their vector representations. This enables us to learn high-quality representations of the products. The experiments demonstrate that BASTEXT can efficiently model millions of baskets, and that it outperforms the state-of-the-art methods in the next product recommendation task. Besides, we will also show that BASTEXT is a strong baseline for keyword-based product search.
\end{abstract}

\maketitle

\section{Introduction}
\label{sec:introduction}
With the rapid development of the internet and online shopping services, modern consumers are able to access a huge amount of products.
During the interaction with the system, consumers leave footprints such as purchase data. Such data is valuable in developing recommender systems that can suggest products that meet the needs of customers.

In this work, we focus on shopping transaction data. A shopping transaction, also known as a \textit{shopping basket}, or a \textit{basket}, is a set of products that a customer buys in a single shopping trip. Such data could help reveal the relationships between products, which are keys to make recommendations in a given context. For example, when a customer is examining a mobile phone case, it is better to recommend him/her other mobile phone cases, or other accessories such as screen protectors. It does not make sense to show a \textit{t-shirt} in that context.

A common approach to shopping basket analysis is association-rules \cite{agarwal1994fast}, which discovers the rules in the form: "\textit{Consumers who buy diapers are likely to buy beer}". However, in a system with a large number of products, many relevant products have never co-occurred in any baskets. The relationships between such products cannot be discovered by association rules. An approach to the context-based recommendation is neighborhood-based methods \cite{sarwar01itembased,Linden2003}, which rely on the similarities between products. A drawback of this approach is that it takes into account only the last product, ignoring the previous ones, which are also valuable for predicting next products. For example, suppose \{\textit{milk}, \textit{sugar}, \textit{egg}\} are in the current shopping basket. Considering all three products is a better indication of buying \textit{flour} rather than considering only \textit{egg}, the last product. Moreover, since both approaches rely on the purchase data, they are not able to model new products. This problem is known as the \textit{cold-start} problem.

Addressing the cold-start using textual contents is well-studied, particularly in recommender systems \cite{Wang:2011:CTM:2020408.2020480,conf_kdd_WangWY15,Li:2017:CVA:3097983.3098077}. These methods are a combination of a text model such as a Variational AutoEncoder \cite{Li:2017:CVA:3097983.3098077} and a matrix factorization-based model \cite{salakhutdinov2008a}. They learn item representations from texts that are useful for predicting the elements of a user-item matrix. However, these models are suitable for modeling long-term preferences of users rather than modeling the relationships between products in shopping baskets.

Recently, neural network-based approaches archive tremendous success in learning text representations \cite{Le:2014:DRS:3044805.3045025,chen2017efficient}.
Though these models are effective in learning text representations, they are not appropriate for understanding shopping baskets. The reason is that the text representations learned by these models can capture the semantic similarities of the texts, but cannot capture the relationships between texts that co-occur in baskets. For example, they cannot identify that \textit{milk} and \textit{flour} often co-occur in baskets because there is no semantic similarity between the titles of these products.

\textbf{This paper:} To address the aforementioned problems, we propose BASTEXT, a novel model for learning product representations from the textual contents, that are useful in explaining the shopping baskets. By learning such representations, BASTEXT enables us to make different types of recommendations such as the products to be added to the current basket, and the products that are often purchased together with a specific product.

Technically, BASKET consists of two \textit{text encoders}, which map the textual contents of the products in a basket and a potential product into fixed-size vector representations. A classifier will identify whether the potential product is relevant to the basket or not, based on their vector representations. This enables us to learn product representations that are strong in identifying which products are likely to be in the same basket, which products are not. Since the basket data is not an obvious dataset for training a classifier, we will present how to form such data from the baskets.


The advantages of the BASTEXT are as follows:
\begin{itemize}
\item It is a scalable model. Since the classifier operates on low-dimensional vector representations, it can model millions of baskets efficiently.
\item It is a flexible model. It allows various types of text encoders to be used. It also enables to use of pre-trained word vectors for learning better representations.
\item It is a multi-purpose model. It can recommend the products in various scenarios. It can also be a strong baseline for keyword-based product search.
\end{itemize}


\section{Related Work}
\label{sec:related_work}
\subsection{Recommender Systems}
Most of the existing recommender systems rely on collaborative filtering (CF) \cite{salakhutdinov2008a,nguyen2018npe,nguyen2015city,nguyen2017hierarchical,nguyen2017collaborative,nguyen2017probabilistic,10.1007/978-3-319-70139-4_20}, which learns user preferences from their prior behaviors such as ratings, purchases, or clicks. One of the most efficient methods for CF is matrix factorization (MF) which models user preferences based on the user-item matrix \cite{salakhutdinov2008a,nguyen2017probabilistic,nguyen2017hierarchical}.
However, MF is strong in identifying users' long-term preferences rather than making recommendations in a given context.

Another line of recommendation is \textit{sequential recommendation} which considers the interactions of users to items as a sequence with an explicit order, e.g., a sequence of clicks. A common approach to this problem is Markov chain-based methods \cite{ShaniMDP,ChenMarkovEmbedding}. Recently, recurrent neural network (RNN) is also used to this problem \cite{Hidasi2015SessionbasedRW}. Note that, our problem is different from sequential recommendation. In shopping basket modeling, there is no explicit order in which the products are added to the baskets. Although a customer adds products to the basket sequentially, the order in which the products are added does not change the nature of the basket.

\subsection{Shopping Basket Analysis}
The most common approach to shopping basket analysis is \textit{association rules}, which discovers the rules in the form: ``\textit{Consumers who buy diapers are likely to buy beer}''. Formally, such rules can be expressed as $B => i$, where $B$ is a set of products and $i$ is a product not contained in $B$. Such rules are useful in making recommendations given the products currently in the basket. However, association rules cannot discover the relationships between products which are relevant but have never co-occurred in the same baskets.,

Another direction of basket analysis is \textit{next basket} recommendations \cite{RendleFPMC,Wang:2015:LHR:2766462.2767694,Yu:2016:DRM:2911451.2914683} which is to suggest a specific customer a whole basket, given his previously shopping baskets. Our work is different.
We are focusing on recommending the next product to add to the current basket rather than recommending the whole basket.


\subsection{Text Representation Learning}


Neural approaches for learning text representations range from simply composition of the word vectors \cite{confnipsMikolovSCCD13,pennington2014glove} to more complicated networks such as doc2vec \cite{Le:2014:DRS:3044805.3045025}; CNN-based approaches \cite{kim2014convolutional,conf/emnlp/WestonCA14}; RNN-base approaches \cite{DBLP:conf/acl/TaiSM15}. Skip-through \cite{Kiros:2015:SV:2969442.2969607} is another model, which learn sentence representations by predicting the surrounding sentences of a given one.

Though they are effective in learning text representations, they may not be appropriate for understanding shopping baskets. The reason is that the goal of these approaches is to learn the text representations that capture the similarities between texts, however, in shopping basket modeling, we need to capture not only the similarities between texts but also the relationships between texts that co-occur in baskets.

\section{BASTEXT: The Shopping Basket Model}
\label{sec:methodology}

\subsection{Notations and Definitions}
Suppose that we have a collection of $T$ shopping baskets. The products in the baskets come from a set of $M$ products, which are denoted by their indices $1, 2, \dots, M$. For each product $i$, there is a text $s_i$ (e.g., \textit{product title}, \textit{product description}, or the set of \textit{tags}) associated with it. We use $\textbf{w}_i$ to denote the input vector of the $i^{th}$ word.
Table \ref{tab:notations} lists the relevant notations used throughout this paper.

\begin{table}[!t]
  \centering
  \caption{The notations used throughout the paper.}
  \label{tab:notations}
  \renewcommand{\arraystretch}{1.2}
  \begin{tabularx}{\linewidth}{c|X}
    Notation&Meaning\\
    \hline
    $T$ & the number of shopping baskets\\
    $M$ & the number of products\\
    $s_i$ & the text associated with the $i^{th}$ product\\
    $B$ & a basket\\
    $\mathbf{w}_i$ & the input vector of the $i^{th}$ word\\
    $K$ & the embedding size\\
    $\mathbf{h}_i$ & the embedding vector of product $i$\\
    $\mathbf{h}'_i$ & the context vector of product $i$\\
    $\overline{\mathbf{h}}'_B$ & the average of the context vectors of the products currently in basket $B$\\
    $\mathcal{D}^+$ & the set of positive examples\\
    $\mathcal{D}^-$ & the set of negative examples\\
    $\mathcal{D}$ & the set of all examples: $\mathcal{D}=\mathcal{D}^+\cup\mathcal{D}^-$\\
    $n$ & the negative sampling ratio
\end{tabularx}
\end{table}


\begin{problem} (\textbf{Next product recommendation}) The task is to recommend the next product to add to the current shopping basket given the products already in the cart.
\end{problem}

\begin{figure*}[!t]
\centering
\includegraphics[scale=1.0]{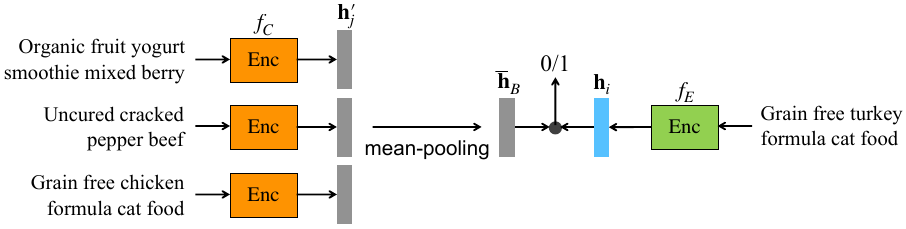}
\caption{The general architecture of BASTEXT framework. The left side is the set of the products already in the current basket. The right side is a potential product.}
\label{fig:basket_architecture}
\end{figure*}

\subsection{Proposed Model}
First, we present the decision process of how a customer chooses a product to add to the current basket. We assume the customer identity is anonymous because many websites allow shopping without account registration. We posit that the customer adds products into the basket sequentially. At each step, the customer chooses one product from the available ones, conditioning on the products already in the basket.

The general architecture is shown in Fig.\ref{fig:basket_architecture}. Given the texts of the products currently in a basket $B$ (the left side) and the text of a potential product $i$ (the right side), the model predicts whether $i$ will be added to the $B$ or not. First, the texts of the products are encoded into fixed-sized vectors. We then apply the mean pooling operation to the vector representations of the products in $B$ to obtain the vector representation of the basket. Second, a classifier identifies whether the potential product $i$ should be added into basket $B$ or not.



Here, we use two text encoders, $f_E$ and $f_C$, for encoding the text $s_i$ of the potential product $i$ and the text $s_j$ of each product $j$ in the basket $B$, respectively. These encoders have the same architecture but with different weights.
\begin{equation}
    \begin{aligned}
    \mathbf{h}_i&=f_E(s_i) \in\mathbb{R}^K, \quad \mathbf{h}'_j=f_C(s_j) \in\mathbb{R}^K\\
    \end{aligned}
\end{equation}

Formally, the probability that $i$ will be added to $B$ is formulated as follows.
\begin{equation}
    \label{eq:continuous_bag_of_texts}
    p(next=i|B)=\sigma\Big(\mathbf{h}^\top_i\overline{\mathbf{h}}'_B\Big)
\end{equation}
where $\overline{\mathbf{h}}'_B=\frac{1}{|B|}\sum_{j\in B} \mathbf{h}'_j$.

Though we can use one encoder, using two encoders has its advantage. If only one encoder is used, each product is represented by one vector, thus can represent only one aspect of the product (e.g., the product's attribute). In contrast, when two encoders are used, each product will be represented by two vectors: the \textit{embedding vector} $\mathbf{h}_i$ and the \textit{context vector} $\mathbf{h}'_i$. The embedding vector identifies the attributes of the product, while the context vector identifies the products that often co-occur with it in the same basket. That is why BASTEXT is able to identify two types of relationships between products: ``similar products'' and ``also-buy products'' as demonstrated in the experiments.



The architecture of a text encoder is a modeling choice. It can be simply the average of the vector representations of its words, a convolutional neural network (CNN) or a recurrent neural network (RNN). In this paper, we implemented two types of text encoders: (1) Mean of vectors (MoV)-based text encoder, and (2) CNN-based text encoder.

\textbf{MoV-based Text Encoder.} The representation of a text is simply the mean of the representations of the words contained in it. In order to introduce non-linearity, we added ReLU after the average layer. The formal specification of the MoV-based Text Encoder is as follows.
\begin{equation}
    f(s) = ReLU\Big(\frac{1}{|s|}\sum_l \mathbf{w}^\top_l\mathbf{W}\Big)
\end{equation}
where $\mathbf{W}$ is the connection matrix of the embedding layer.

Though this network is simple, it has two advantages: (1) it is very efficient in computational cost, (2) it can be used when there is no explicit order of the words in a text, e.g., when the text is a set of tags associated with a product.

\textbf{CNN-based Text Encoder.} Although the MoV-based Text Encoder is simple and efficient, it ignores the order of words in sentences. In order to verify the effectiveness of the word order, we implement a CNN-based Text Encoder, which can take into account the order of the words in a text. Though using a recurrent neural network (RNN) is common in modeling sentence, we use CNN due to its efficiency in computational cost. We use the CNN architecture proposed in \cite{kim2014convolutional}.

\subsection{Forming Training Data}
\label{sec:training_data_forming}
We present how to form the training data from a collection of shopping baskets. Each training example is a tuple: $\big \langle(B, i), L\big \rangle$ where $B$ is the set of products currently in the basket, $i$ is a potential product, and $L$ is the label $+$ or $-$ to indicate whether $i$ was chosen or not.

\textbf{Positive examples.} For each basket, we pick each product in turn and use as the potential product, the remaining products are interpreted as the products currently in the basket. By this way, we obtain $m_B$ positive examples, where $m_B$ is the number of products in the basket $B$.

\textbf{Negative examples.} Since the negative examples are not available, we obtain them by \textit{negative sampling}). Here, we use the uniform sampling method. Other strategies are left for future work. For each positive example $\big \langle(B, i), +\big \rangle$, we randomly pick a product $j$ which is not in $B$ to form a negative example $\big \langle(B, j), -\big \rangle$. For each positive example, we repeat this procedure $n$ times to obtain $n$ negative examples.

\subsection{Parameter Learning}
After forming the training data, we have a set of examples $\mathcal{D}$, where each example is in the form $\big \langle(B, i), L\big \rangle$, where $L$ is $-$ or $+$. The objective function is the negative log likelihood over all examples in the training set which is formulated as:
\begin{align}
\label{eq:objective_function}
\mathcal{L}(\Theta) & =\sum_{(B, i)\in {\mathcal{D}^+}}\log\mu_{B,i}-\sum_{(B, i)\in {\mathcal{D}^-}}(1-\log\mu_{B,i})
\end{align}
where $\mu_{B,i}=\sigma(\mathbf{h}_i^\top\overline{\mathbf{h}}'_B)$.

Training the BASTEXT model can be efficiently performed by back-propagation using stochastic gradient descent with mini-batch. In the experiments, we use Adam \cite{DBLP:journals/corr/KingmaB14}. We do not perform negative sampling in advance. Instead, we use negative sampling at each mini-batch for obtaining diverse negative examples.

\section{Experiments}
\label{sec:experiments}
\subsection{Datasets}
We use two public datasets of varying sizes:
\begin{itemize}
  \item \textbf{OnlineRetail} \cite{onlineretail}: this dataset contains about 20,000 shopping baskets. The average number of products in a basket is $26.7$. The average length of the product descriptions is $4.3$ words.
  \item \textbf{Instacart}\footnote{https://www.instacart.com/datasets/grocery-shopping-2017}: this dataset contains 3.2 million orders, where the average number of products per order is $10.6$. Each product is associated with a product title whose average length is $4.7$ words.
\end{itemize}

\subsection{Experimental Setup}
We randomly split the baskets into three sets: \textit{training baskets}, \textit{validation basket} and \textit{testing baskets}, with proportions $85\%, 5\%, 10\%$. Then we form the \textit{training set}, \textit{validation set} and \textit{test set} in two ways: \textbf{warm-start} and \textbf{cold-start}. The details about data splitting is given in Table \ref{tab:warm_start_splitting} and Table \ref{tab:cold_start_splitting}.
\begin{table}[!t]
\caption{The statistical information of the datasets}
\label{tab:data_info}
\renewcommand{\arraystretch}{1.2}
\centering
  \subcaption{Warm-start splitting}
  \label{tab:warm_start_splitting}
  \begin{tabular}{l|c|c}
    \hline
    Data & OnlineRetail & Instacart\\
    \hline
    \hline
    \# training baskets & 17K & 2.7M\\
    \# validation baskets & 1K & 159K\\
    \# test baskets & 1.9K & 318K\\
    \# test cases & 51K & 3.3M\\
    \hline
  \end{tabular}
  \bigskip\smallskip
  \subcaption{Cold-start splitting}
  \label{tab:cold_start_splitting}
  \begin{tabular}{l|c|c}
    \hline
    Data & OnlineRetail & Instacart\\
    \hline
    \hline
    \# training baskets & 16K & 2.3M\\
    \# validation baskets & 988 & 138K\\
    \# test baskets & 1.7K & 312K\\
    \# test cases & 13.6K & 2.3M\\
    \hline
  \end{tabular}
\end{table}

\textbf{Warm-start.} In this setting, we make sure that every product in the test set appears in the training set. To do that, we remove from the test baskets the products that do not appear in the training baskets before forming the training set, the validation set, and the test set.

\textbf{Cold-start.} In this setting, we make sure that every product in the test set is absent from the training set. We randomly pick $10\%$ products from the test baskets and call these products \textit{test products}. We remove these products from train baskets. Then we form the test cases in which the potential products come from the \textit{test products}. The validation set and training set are formed similarly with the warm-start setting.

\textbf{Evaluation.} For each basket in the test set, we predict the relevant scores for all the remaining products and rank these products according to their relevance scores. We then pick $N$ products that have highest scores to form a recommendation list. We use common rank-based metrics: Recall@$N$, and MRR@$N$ (mean reciprocal rank) for evaluating the models.



\begin{table*}[!t]
\caption{Recall and MRR for next product recommendation (warm-start setting). Here, we fixed the embedding size $K=64$ and negative sampling ratio $n=8$.}
\label{tab:next_product_warm_start}
\renewcommand{\arraystretch}{1.2}
\centering
\begin{subtable}[t]{0.48\textwidth}
  \caption{OnlineRetail data}
  \label{tab:online_retail_warm_start}
  \centering
  \footnotesize
\begin{tabular}{l|c|c|c}
    \hline
    Methods & Re@10 & Re@20 & MRR@20\\
    \hline
    \hline
    POP & 0.0828 & 0.1212 & 0.0652\\
    ItemKNN & 0.1923 & 0.2511 & 0.1765\\
    prod2vec & 0.2007 & 0.2632 & 0.1876\\
    doc2vec & 0.1773 & 0.2332 & 0.1521\\
    \hline
    BASTEXT-Avg (our) & {0.2181} & {0.2854} & {0.1943}\\
    BASTEXT-Avg+ (our) & {0.2275} & {0.2942} & {0.2132}\\
    BASTEXT-Conv (our) & {0.2212} & {0.2897}& {0.1998}\\
    BASTEXT-Conv+ (our)  & \underline{0.2378} &  \underline{0.3096} & \underline{0.2251}\\
    \hline
  \end{tabular}
\end{subtable}
\bigskip
\bigskip\smallskip
\hfill
\begin{subtable}[t]{0.48\textwidth}
  \caption{Instacart data}
  \label{tab:instacart_warm_start}
  \centering
  \footnotesize
  \begin{tabular}{l|c|c|c}
    \hline
    Methods & Re@10 & Re@20 & MRR@20\\
    \hline
    \hline
    POP & 0.0124 & 0.0153 & 0.0102\\
    ItemKNN & 0.1065 & 0.1507 & 0.0985\\
    prod2vec & 0.1251 & 0.1623 & 0.1048\\
    doc2vec & 0.0912 & 0.1215 & 0.0981\\
    \hline
    BASTEXT-Avg (our) & {0.1527} & {0.1932} & {0.1329}\\
    BASTEXT-Avg+ (our) & {0.1631} & {0.2013} & {0.1521}\\
    BASTEXT-Conv (our) & {0.1578} & {0.1965}& {0.1401}\\
    BASTEXT-Conv+ (our)  & \underline{0.1698} &  \underline{0.2102} & \underline{0.1598}\\
    \hline
  \end{tabular}
\end{subtable}
\end{table*}

\begin{table*}[!t]
\caption{Recall and MRR for next product prediction (cold-start setting). Here, we fixed the embedding size $K=64$ and negative sampling ratio $n=8$.}
\label{tab:next_product_cold_start}
\renewcommand{\arraystretch}{1.2}
\centering
\begin{subtable}[t]{0.48\textwidth}
  \caption{OnlineRetail data}
  \label{tab:online_retail_cold_start}
  \centering
  \footnotesize
\begin{tabular}{l|c|c|c}
    \hline
    Methods & Re@10 & Re@20 & MRR@20\\
    \hline
    \hline
    doc2vec & 0.1768 & 0.2315 & 0.1532\\
    \hline
    BASTEXT-Avg (our) & {0.1823} & {0.2378} & {0.1628}\\
    BASTEXT-Avg+ (our) & {0.1861} & {0.2432} & {0.1679}\\
    BASTEXT-Conv (our) & {0.1842} & {0.2397}& {0.1642}\\
    BASTEXT-Conv+ (our)  & \underline{0.1908} &  \underline{0.2483} & \underline{0.1733}\\
    \hline
  \end{tabular}
\end{subtable}
\bigskip\smallskip
\hfill
\begin{subtable}[t]{0.48\textwidth}
  \caption{Instacart data}
  \label{tab:instacart_cold_start}
  \centering
  \footnotesize
  \begin{tabular}{l|c|c|c}
    \hline
    Methods & Re@10 & Re@20 & MRR@20\\
    \hline
    \hline
    doc2vec & 0.0916 & 0.1208 & 0.0977\\
    \hline
    BASTEXT-Avg (our) & {0.1021} & {0.1297} & {0.1048}\\
    BASTEXT-Avg+ (our) & {0.1098} & {0.1385} & {0.1195}\\
    BASTEXT-Conv (our) & {0.1075} & {0.1342}& {0.1127}\\
    BASTEXT-Conv+ (our)  & \underline{0.1127} &  \underline{0.1428} & \underline{0.1249}\\
    \hline
  \end{tabular}
\end{subtable}
\end{table*}

\textbf{Competing Methods.} In evaluating the predictive performance, we compare the following methods (including ours). We do not compare with MF-based methods because these methods are not appropriate for modeling shopping baskets.
\begin{itemize}
  \item \textbf{POP} (popular products): this model recommends the most popular products in the training set. Though POP is simple, it is often a strong baseline in certain domains.
  \item \textbf{ItemKNN} \cite{Linden2003}: this model is based on the co-occurrences of products in the baskets. This is one of the most common \textit{item-to-item} recommendation in the form ``users who bought $X$ also bought $Y$''.
  \item \textbf{prod2vec} \cite{Grbovic:2015:EYI:2783258.2788627}: a word2vec version for learning the product representations by corresponding a basket as a sentence and a product in the basket as a word. A basket's representation is calculated as the mean of the products contained in the basket. Given a basket, we compute the \textit{cosine similarities} between its representation and all potential products, and pick top $N$-similar products.
  \item \textbf{doc2vec} \cite{Le:2014:DRS:3044805.3045025}: a model for learning text representations. We apply doc2vec to obtain the product representations from their titles. A basket's representation is calculated as the mean of the products contained in it. Given a basket, we calculate the \textit{cosine similarities} between the basket's representation and all potential products, and pick top-$N$ similar products.
  \item \textbf{BASTEXT-Avg} (our): the BASTEXT model where the MoV-based text encoders are used for learning text representations. The word input vectors are one-hot vectors.
  \item \textbf{BASTEXT-Avg+} (our): the BASTEXT-Avg model where the input vector for each word is the pre-trained word vector \cite{pennington2014glove}.
  \item \textbf{BASTEXT-Conv} (our): the BASTEXT model where the CNN-based text encoder is used for learning the representations of texts. The input vector for each word is its one-hot-vector.
  \item \textbf{BASTEXT-Conv+} (our): the BASKET-Conv model where the input vector for each word is the pre-trained word vector \cite{pennington2014glove}.
\end{itemize}

\subsection{Implementation Detail}
All BASTEXT variants are trained by optimizing the binary cross-entropy loss in Eq.\ref{eq:objective_function}. We use dropout \cite{Srivastava2014} for hidden layers to avoid over-fitting. To speed up the training process, we exploit the power of GPU. In dividing the training data into mini-batch, we choose the mini-batch sizes that fit the GPU's memory. The mini-batch size is 10,000 for OnlineRetail data and 5,000 for Instacart data.

\subsection{Comparison over Baselines}
Table \ref{tab:next_product_warm_start} and Table \ref{tab:next_product_cold_start} show the performances of the next product prediction task. We can see that all variants of BASTEXT significantly outperform the other methods. Besides, we have the following observations.

As expected, POP does not achieve good performance because it cannot capture the context of the shopping trips. Therefore, it is easily beaten by ItemKNN and prod2vec. We also can see that ItemKNN and prod2vec outperform doc2vec, which uses content only, indicating that the basket data is more valuable than the contents in capturing the shopping behaviors.

In the warm-start setting, BASTEXT-Avg significantly outperforms prod2vec ($8.4\%$ and $19\%$ for OnlineRetail and Instacart, respectively). It indicates that introducing textual contents will significantly improve the performances.

In the cold-start setting, only doc2vec and the variants of BASTEXT work. The performances of doc2vec are almost the same as the warm-start setting because it uses content only. We can see that BASTEXT-Avg performs slightly better than doc2vec. This indicates that jointly training the texts with purchase data will improve the performance.

\textbf{Impact of the Text Encoder model.} From Table \ref{tab:next_product_warm_start} and \ref{tab:next_product_cold_start} we can observe that BASTEXT-Conv and BASTEXT-Conv+ slightly perform better than their counterparts (BASTEXT-Avg and BASTEXT-Avg+). This indicates that taking into account the order of words in texts can improve the representations. However, such minor improvements suggest that using the MoV-based text encoder is also effective, given that its complexity is cheaper than a CNN.

\textbf{Impact of the pre-trained word vectors.} One advantage of BASTEXT is that it can use the pre-trained word embedding vectors as its input. Thus, we can study the impact of using pre-trained word embedding vectors on the performance of BASTEXT. Here, we use the pre-trained vectors of GloVe \cite{pennington2014glove}.

Table \ref{tab:next_product_warm_start} and Table \ref{tab:next_product_cold_start} show marginal improvements of BASTEXT-Avg+ and BASTEXT-Conv+ over their counterparts (BASTEXT-Avg and BASTEXT-Conv).
It is because the product titles are very short, therefore, are poor at capturing the semantic meaning of the texts. Using pre-trained word vectors will improve the representations of short texts.

\begin{figure*}[!t]
\begin{minipage}[b]{.48\textwidth}
\centering
\includegraphics[width=0.9\textwidth]{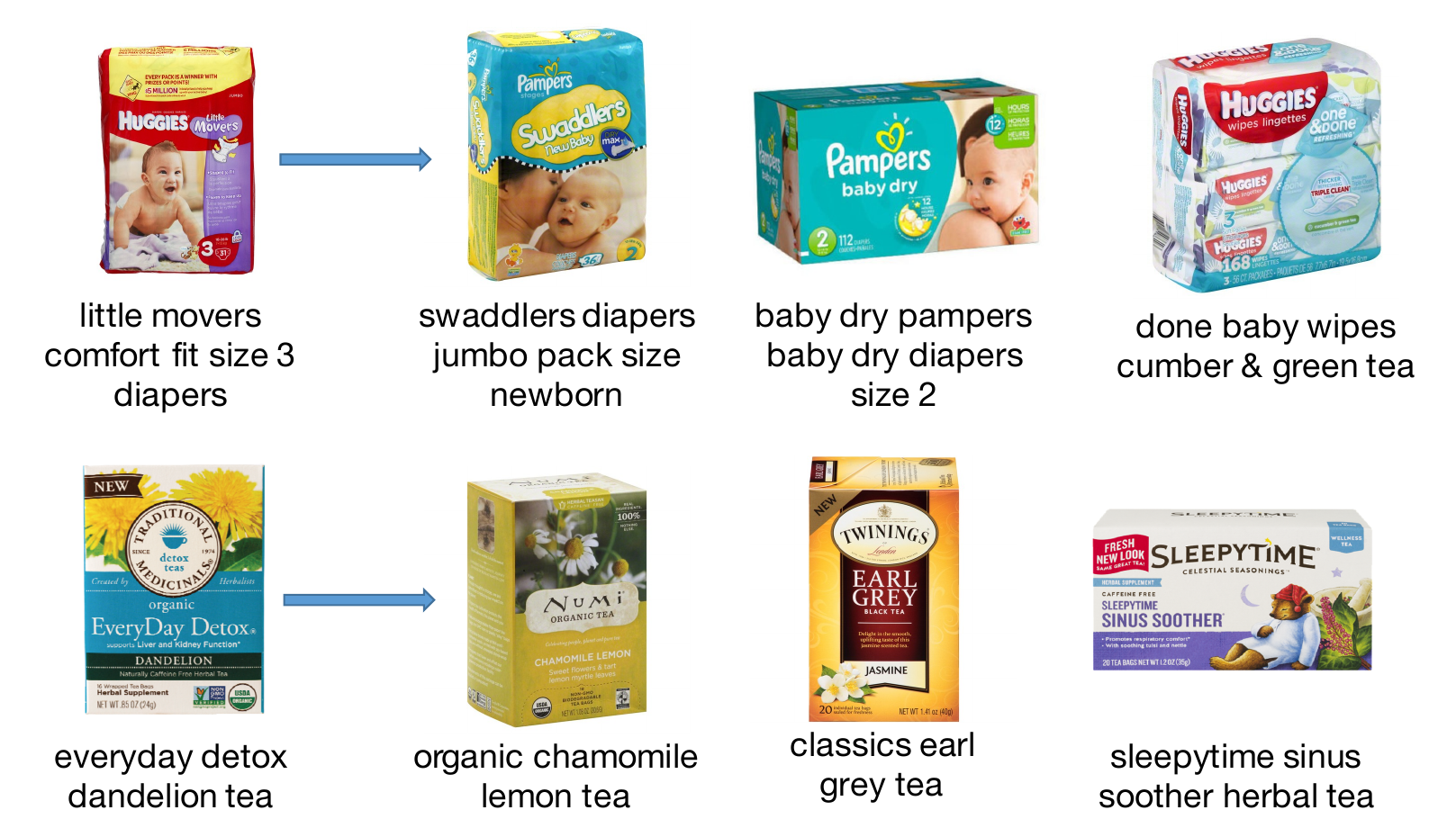}
\small
\caption{Similar product recommendation. For each row, the product in the left side is a ``query'' product, following by its top-3 similar products.}
\label{fig:similar_product_instacart}
\end{minipage}
\hfill
\begin{minipage}[b]{.48\textwidth}
\centering
\includegraphics[width=0.9\textwidth]{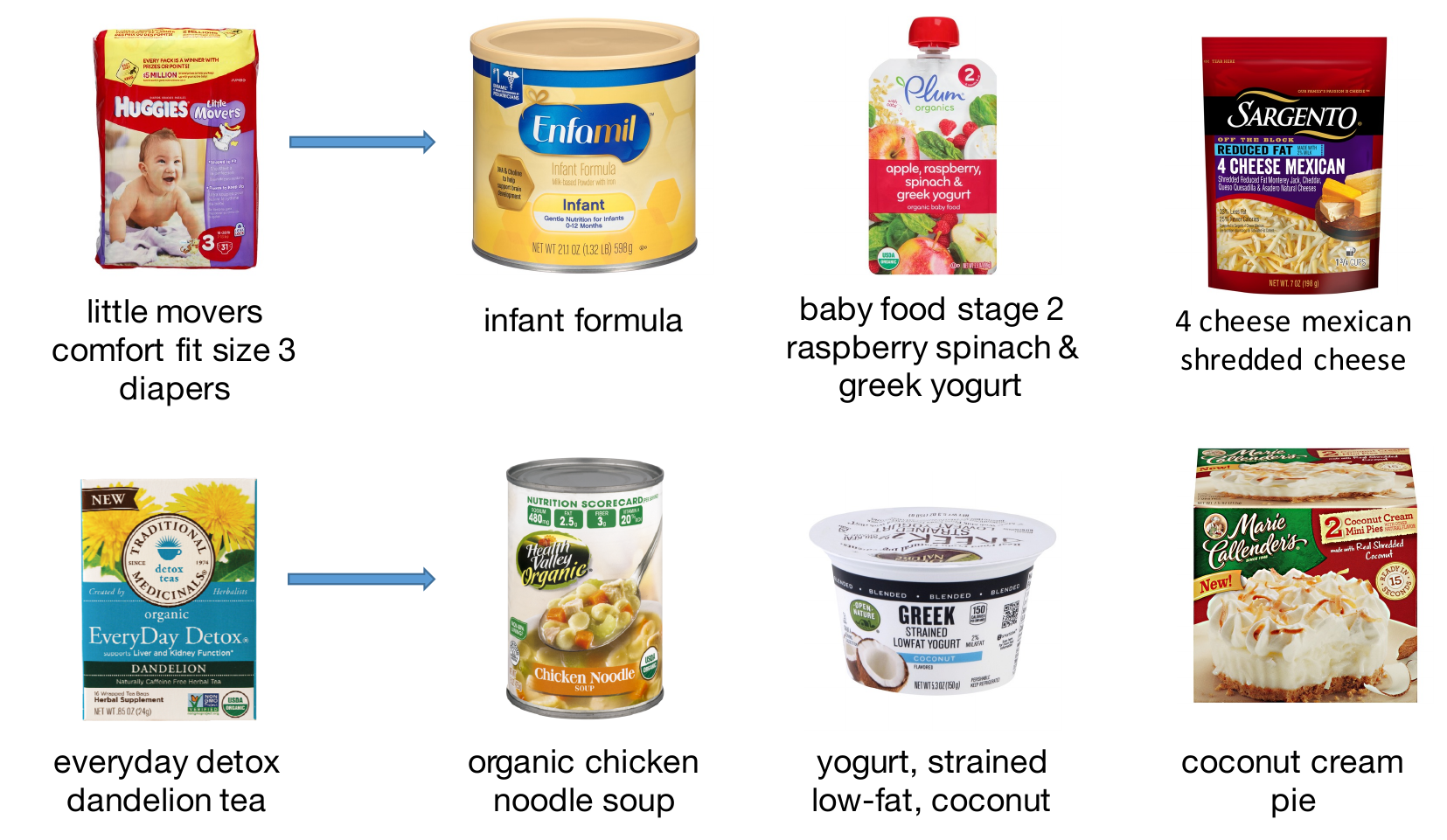}
\small
\caption{Also-buy product recommendation. For each row, the product in the left side is the ``query'' product, following top-3 products that are often co-purchased with it.}
\label{fig:also_buy_product_instacart}
\end{minipage}
\end{figure*}

\subsection{Product-based Recommendation}
Making recommendations in the context of a specific product is a typical scenario. Here, we consider two kinds of such recommendations: \textit{similar product} and \textit{also-buy product} recommendations.

\textbf{Similar product recommendation.} This is useful when a customer is examining a product. For example, if the customer is examining a skirt, it makes sense to show her some other skirts so that she can compare before deciding.

The similarity between two products $i$ and $j$ is defined as the \textit{cosine similarity} between their embedding vectors:

\begin{equation}
    \label{eq:similar_products}
    sim(i,j)=cosine(\mathbf{h}_i, \mathbf{h}_j)
\end{equation}

Fig.\ref{fig:similar_product_instacart} shows some examples of similar product recommendations. In each row, the most left is a ``query'' product, and the next three products are its top-3 similar products, calculated by Eq.\ref{eq:similar_products}.


\textbf{Also-buy Product Recommendation.} This is to recommend products that are frequently purchased together with a specific product. This scenario is useful when a customer has added a product to his shopping basket.

Given two products $i$ and $j$, we compute how likely $i$ is bought given that the customer has already bought $j$ as the \textit{inner product} of the context vector $\mathbf{h}_i$ and $\mathbf{h}'_j$:
\begin{equation}
    \label{eq:also_buy_products}
    Also\_buy(i,j)=\mathbf{h}^\top_i\mathbf{h}'_j
\end{equation}

Fig.\ref{fig:also_buy_product_instacart} shows some examples of also-buy product recommendations. In each row, the most left is a ``query'' product, and the next three products are its top-3 ``also-buy''products, calculated by Eq.\ref{eq:also_buy_products}.


\subsection{Effectiveness of the Representations}
We study how well BASTEXT captures the semantics behind the products by performing two tasks: \textit{product search} and \textit{product category classification}.

\textbf{Product search.} Given a query $s$ in the form of keywords, the task is to retrieve the products relevant to the query. We compare BASTEXT with doc2vec \cite{Le:2014:DRS:3044805.3045025}.

First, we infer the vector representations of the query by two models BASTEXT-Avg and doc2vec. For BASTEXT, we use the text encoder $f_E$. We then compute the cosine similarity query's vector representation with the embedding vector of every product in the dataset. Top-5 similar products are reported in Table \ref{tab:keyword_based_search}.

We observe that BASTEXT retrieves more relevant products than doc2vec does. Especially, for the second query, ``\textit{natural herb cough drops}'', BASTEXT can return relevant products while doc2vec completely misunderstands the query. We found that the keywords of this query rarely appear in the titles, therefore, doc2vec cannot learn good representations. BASTEXT, in contrast, can learn effective representations by leveraging the basket data. This experiment suggests that BASTEXT can be a potential baseline for product search, especially when the product titles are short.
\begin{table*}[!t]
  \centering\arraybackslash
  \caption{Product search results on Instacart data. The top line are the query (in the boldface font). Below the query are top five answers by BASTEXT-Avg and word2vec, respectively. Inside the braces $()$ are the categories of the returned products. Underlined words are words appear in the query.}
  \label{tab:keyword_based_search}
  \renewcommand{\arraystretch}{1.2}
  \small
  \begin{tabularx}{1.0\linewidth}{l|l|l}
    query & \textbf{organic tea}&\textbf{natural herb cough drops}\\
    \hline
    \multirow{5}{*}{BASTEXT} & \underline{organic} honeybush \underline{tea} (tea)& \underline{cough} drop (cold flu allergy)\\
    &\underline{organic} chamomile lemon \underline{tea} (tea) & honey/lemon \underline{cough} drops (cold flu allergy)\\
    &\underline{organic} white rose white \underline{tea} (tea) & defense vitamin c, cold flu allergy (cold flu allergy)\\
    &chinese breakfast black \underline{tea} (tea) & \underline{natural} throat \underline{drops} honey \& pomegranate (cold flu allergy)\\
    \hline
    \multirow{5}{*}{doc2vec} & \underline{organic} english breakfast black \underline{tea} (tea) & ultra thin crust cheese lovers pizza (frozen pizza)\\
    &lemon sweet tea iced \underline{tea} mix (tea) & homemade pizza sauce (pasta sauce)\\
    &bags \underline{organic} turmeric ginger green \underline{tea} (tea) & authentic deep dish sausage pizza (frozen pizza)\\
    &half sweet \underline{tea} pink lemonade (tea) & colby jack cheese (packaged cheese)\\
\end{tabularx}
\end{table*}

\begin{figure}[!t]
  \centering
  \includegraphics[width=1.0\linewidth]{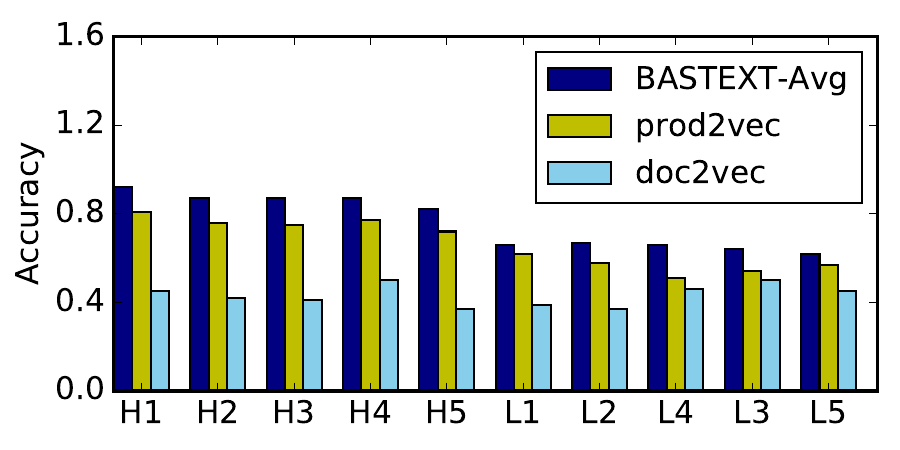}
  \caption{Performance of product category classification on Instacart.}
  \label{fig:classification_result_svm}
\end{figure}

\textbf{Category classification.} We additionally study the effectiveness of the product representations of BASTEXT, prod2Vec, and doc2vec, by performing category classification task on Instacart dataset. To do so, we use the embedding vectors of the products learned by these models as the feature vectors and use Support Vector Machine as the classifier. We perform 5-fold cross-validation and report the classification accuracies. The products used in the test are from two groups. The first group contains 5 \text{most active} categories (categories that are most frequently purchased): (H1) Produce, (H2) Dairy eggs, (H3) Snacks, (H4) Beverages, (H5) Frozen. The second group contains 5 \textit{least active} categories (categories that are less frequently purchased): (L1) Personal care, (L2) Babies, (L3). (L4) Alcohol, (L5) Pets.

The result is shown in Fig.\ref{fig:classification_result_svm}. The accuracies of doc2vec are almost the same across the categories. This is expected because doc2vec uses only the textual content. In the \textit{least active} categories (L1-L5), BASTEXT and prod2vec perform better than doc2vec, however, the differences are not big. The differences between BASTEXT and prod2vec with doc2vec become larger in the \textit{most active} categories (H1-H5), indicating the important role of purchase data in the performance. In the \textit{most active} categories, BASTEXT is still better than prod2vec. This implies that introducing the texts will improve the effectiveness of the representations.

\subsection{Hyper-parameter Sensitivity}
In this section, we study the impact of hyper-parameters to the performance of the next product prediction task.

\begin{figure}[t]
\centering
\begin{subfigure}{0.49\linewidth}
  \centering
  \includegraphics[width=1.0\linewidth]{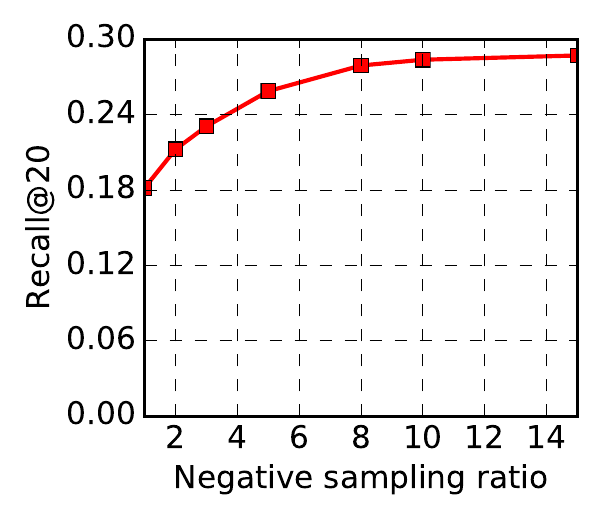}
  \subcaption{Re@20 for OnlineRetail}
  \label{fig:neg_ratio_recall20_online_retail}
\end{subfigure}
\begin{subfigure}{0.49\linewidth}
  \centering
  \includegraphics[width=1.0\linewidth]{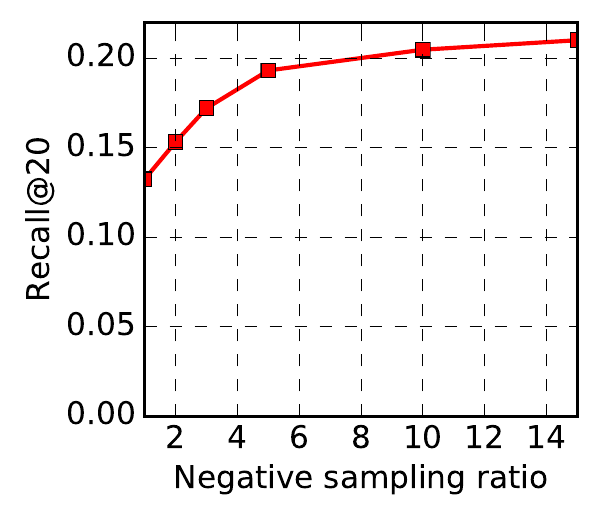}
  \subcaption{Re@20 for Instacart}
  \label{fig:neg_ratio_recall20_instacart}
\end{subfigure}
\caption{Impact of the negative sampling ratio. Here we use BASTEXT-Avg with embedding size $K=64$.}
\label{fig:neg_ratio_impact}
\end{figure}

\textbf{Impact of the Negative Sampling Ratio.}
Fig.\ref{fig:neg_ratio_impact} shows the BASTEXT-Avg's performances of the next product prediction with different negative sampling ratio $n$. We observe that, the Recall@20 increases when the negative ration $n$ increases, until a certain value of $n$ (around 8 to 10) before getting stable. Therefore, we do not need to sample more negative examples than this value of $n$.

\begin{figure}[t]
\begin{subfigure}{0.49\linewidth}
  \centering
  \includegraphics[width=1.0\linewidth]{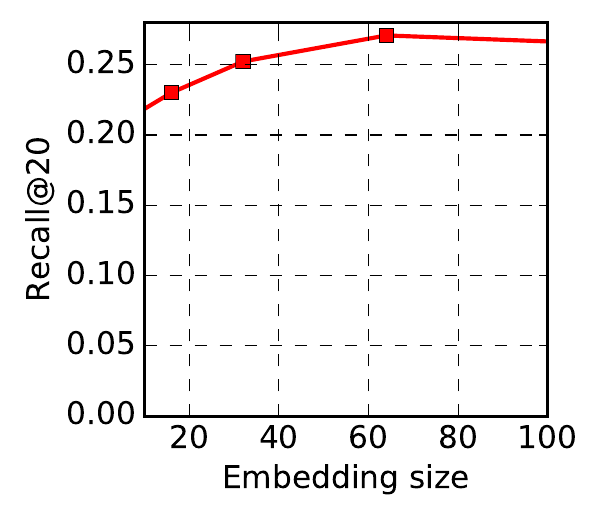}
  \subcaption{Re@20 for OnlineRetail}
  \label{fig:emb_size_recall20_online_retail}
\end{subfigure}
\begin{subfigure}{0.49\linewidth}
  \centering
  \includegraphics[width=1.0\linewidth]{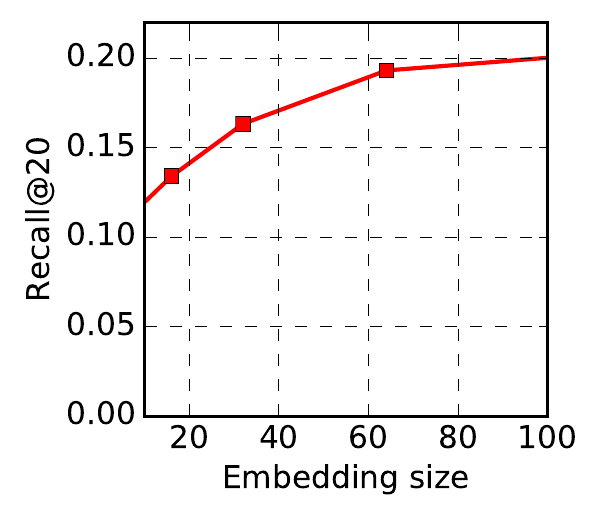}
  \subcaption{Re@20 for Instacart}
  \label{fig:emb_size_recall20_instacart}
\end{subfigure}
\caption{Recall@20 with different embedding sizes. Here we use BASTEXT-Avg with negative sampling ratio $n=8$.}
\label{fig:emb_size}
\end{figure}

\textbf{Impact of the Embedding Size.} Fig.\ref{fig:emb_size} shows the BASTEXT-Avg's performances of the next product prediction with different embedding size $K$. We can observe that the performances increase when $K$ increases, until a threshold of $K$ (around 64). Then, the performances decrease (OnlineRetail data) or do not significantly increase (Instacart data). These observations suggest that the embedding size around $K=64$ will have a balance between the performances and the computational complexity.

\section{Conclusions}
\label{sec:conclusions}
We introduced BASTEXT, a model of texts and shopping basket data. BASTEXT uses the texts for addressing the cold-start problem and uses the basket data for improving the performance of text representations. The experiments show that BASTEXT is effective in various tasks: the next product recommendation, similar product recommendation, also-buy product recommendation, and product search.

There are several directions for the future work. One is to use data such as click data, in addition to the purchasing data. Another direction is to use other auxiliary data such as product images or the user reviews in modeling the products.

\bibliographystyle{ACM-Reference-Format}
\bibliography{bastext2019}

\end{document}